\documentclass[english,english,noshowpacs, amssymb,twocolumn,aps,nofootinbib]{revtex4}
\usepackage[T1]{fontenc}
\usepackage[latin9]{inputenc}
\usepackage{color}
\usepackage{mathtools}
\usepackage{textcomp}
\usepackage{amsmath}
\usepackage{esint}
\usepackage{url}
\usepackage{natbib}
\usepackage{graphicx}
\usepackage[export]{adjustbox}

\makeatletter

\newcommand{\lyxmathsym}[1]{\ifmmode\begingroup\def\b@ld{bold}
  \text{\ifx\math@version\b@ld\bfseries\fi#1}\endgroup\else#1\fi}

\@ifundefined{textcolor}{}
{%
 \definecolor{BLACK}{gray}{0}
 \definecolor{WHITE}{gray}{1}
 \definecolor{RED}{rgb}{1,0,0}
 \definecolor{GREEN}{rgb}{0,1,0}
 \definecolor{BLUE}{rgb}{0,0,1}
 \definecolor{CYAN}{cmyk}{1,0,0,0}
 \definecolor{MAGENTA}{cmyk}{0,1,0,0}
 \definecolor{YELLOW}{cmyk}{0,0,1,0}
}
\makeatother

\usepackage{babel}
\begin{document}

\title{Observation of Spatial Quantum Correlations in the Macroscopic Regime}

\author{Ashok Kumar{\footnote {ashok@ou.edu}}, Hayden Nunley, and  A. M. Marino{\footnote {marino@ou.edu}}}
\affiliation{Homer L. Dodge Department of Physics and Astronomy, The University of Oklahoma, Norman, Oklahoma 73019, USA}

\begin{abstract}
Spatial quantum correlations in the transverse degree of freedom promise to enhance optical resolution, image detection, and quantum communications through parallel quantum information encoding. In particular, the ability to observe these spatial quantum correlations in a single shot will enable such enhancements in applications that require real time imaging, such as particle tracking and in-situ imaging of atomic systems. Here, we report on measurements in the far-field that show spatial quantum correlations in single images of bright twin-beams with $10^8$ photons in a 1~$\mu$s pulse using an electron-multiplying charge-coupled device camera. A four-wave mixing process in hot rubidium atoms is used to generate narrowband-bright pulsed twin-beams of light. Owing to momentum conservation in this process, the twin-beams are momentum correlated, which leads to spatial quantum correlations in the far field. We show around 2~dB of spatial quantum noise reduction with respect to the shot noise limit. The spatial squeezing is present over a large range of total number of photons in the pulsed twin-beams.
\\
\\PACS numbers: 42.50.Dv, 42.50.Ar, 03.67.-a
\end{abstract}

\maketitle

Under certain conditions, the quantum fluctuations in beams of light can be reduced below the shot noise limit (SNL) not only in the temporal domain, but also in the transverse spatial degree of freedom~\cite{RMP}. To date, most of the attention has focused on the study of quantum noise reduction, or squeezing, in the temporal domain~\cite{{Slusher,Fabre,Pkumar,Treps,Lett1,OPO10dB,LettOE,Leuchs1,Glorieux,Bharath,Leuchs2}}. Nevertheless, many areas in quantum optics, specifically quantum metrology and  quantum imaging, could greatly benefit from the study of the quantum correlations directly in the spatial domain~\cite{{Treps1,Kolobov,Genovese}}. This would make it possible to take advantage of the spatial quantum properties of light, such as spatial squeezing, for enhanced spatial resolution and sub shot noise imaging \cite{Kolobov}.

With this in mind, a few groups have recently experimentally demonstrated sub-shot noise spatial correlations using an electron-multiplying charge-coupled device (EMCCD) camera in photon pairs generated through spontaneous parametric down conversion (SPDC)~\cite{shot1,shot2,shot3}. As a proof of principle of a potential application of spatial quantum correlations in quantum imaging, Brida~\textit{et al.}~\cite{Brida} imaged a weak absorbing object with significantly higher signal-to-noise ratio than what is possible with classical light. However, the intensity levels were limited by the source and are orders of magnitude lower than what is used in standard imaging techniques.

While these initial experiments provided an indication that spatial quantum correlations can lead to significant enhancements, many applications in quantum imaging and quantum metrology require real time imaging and, as such, require the ability to observe the spatial quantum correlations in a single shot with a controllable and macroscopic number of photons.  In addition, the use of such a large number of photons can lead to a more significant sensitivity enhancement due to the $\sqrt{N}$ scaling of the signal-to-noise ratio, where $N$ is the number of photons. Thus, if one is able to increase $N$ while preserving the spatial quantum correlations, a significant advantage can be obtained. Here we show that it is possible to observe spatial quantum correlations in the far field with bright twin-beams captured by an EMCCD. We observe such quantum correlations in single images obtained with bright pulsed twin-beams with $\sim10^8$ photons in 1~$\mu$s pulses, which correspond to a photon flux of $10^{14}$ photon pairs per second. This makes bright twin-beams a useful candidate for quantum imaging and quantum metrology  applications that require in-situ real time imaging, such as particle tracking in biological samples~\cite{Bachor}, Bose-Einstein condensates~\cite{BEC-imaging,BEC}, and trapped single atoms~\cite{atom-imaging}.

As a source of bright twin-beams of light, we use a four-wave mixing (FWM) process in a double-$\Lambda$ configuration in an atomic vapor. In recent years, twin-beams have gained considerable attention due to their applications in quantum information, quantum computing, and quantum metrology~\cite{{RMP,Lukin,CV-RMP,Lett2,Lett3,Lett4,Lett5,Lett6,Howell}}. While some of these previous experiments have characterized the spatial quantum correlations of the twin beams, the measurements have always been performed in the temporal domain by selecting different spatial regions through either amplitude masks or homodyning with spatially structured local oscillators~\cite{Lett2,MarinoCoh}.  In order to fully take advantage of the large degree of spatial quantum correlations that can be generated with this source, it is necessary to extend the measurement techniques to the spatial domain.  Here, we measure, for the first time, the presence of spatial intensity-difference squeezing with bright twin-beams directly in the spatial domain with an EMCCD camera.  Thus, this work paves the way for the implementation of quantum imaging in the macroscopic regime.

There are some unique advantages of using bright twin-beams generated through FWM over the faint beams obtained with SPDC. First, the photon pairs generated by FWM have narrow bandwidths (in the MHz regime, even when working with hot atoms)~\cite{Bharath,Lett6,Howell}, therefore they are useful for atom-light interaction-based quantum protocols~\cite{DLCZ}. Second, the FWM process offers large gains even in a single pass configuration unlike SPDC~\cite{Lett1,SPDC-efficiency}. As a result, FWM can produce bright quantum correlated beams of light without a cavity~\cite{OPO10dB}. This makes it possible to preserve the multi-spatial mode nature of the bright twin-beams.

A schematic of our experimental setup and the energy level scheme used for the FWM is shown in Fig.~1. A strong laser beam (2.4~W of power) from a CW Ti-Sapphire laser is used as a pump for the FWM process. The frequency of the laser is locked 1~GHz away from the atomic hyperfine transition F=2 to F$^{\prime}$=3 of the $^{85}$Rb D$_1$ line (wavelength $\sim795$~nm) through a saturation absorption spectroscopy setup. A weak laser beam (power $\sim 70$~$\mu W$) derived from the same laser and frequency down-shifted by 3.04~GHz with an acousto-optic modulator (AOM) acts as the input probe beam for the FWM process. This leads to a 4~MHz detuning from the two-photon transition between the ground states $F=2$ and $F=3$ of $^{85}$Rb. Orthogonally linearly polarized pump and probe beams with $1/e^2$ waist diameters of 4.5~mm and 0.2~mm respectively, interact at an angle of $\sim 0.5^{\circ}$ at the center of a 12~mm long $^{85}$Rb vapor cell heated to 113$^{\circ}$C.  As a result of the FWM in the double-$\Lambda$ configuration, shown in the inset of Fig.~1, two pump photons are absorbed and quantum correlated probe and conjugate photons are generated~\cite{Lett1}. After the cell, most of the pump beam is filtered with a polarization filter.

\begin{figure}[hbt]
\centering
\includegraphics[width=\columnwidth]{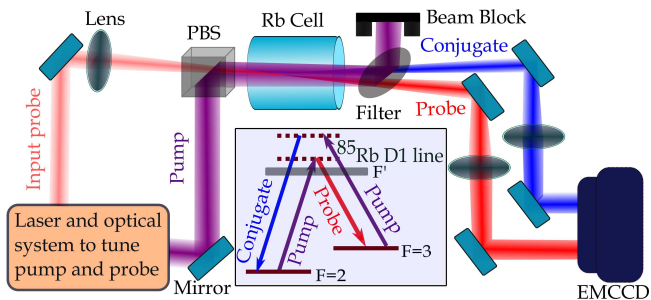}
\caption{Schematic of experimental setup for the four-wave mixing and characterization of spatial quantum correlations in the far-field present in bright twin-light-beams with an EMCCD camera. PBS: polarizing beam splitter. The inset shows the double-$\Lambda$ energy level scheme for the FWM.}
\label{fig:Fig1}
\end{figure}

The output probe and conjugate beams are passed through two separate 50~cm focal lengths lenses in an \emph{f}-to-\emph{f} configuration to obtain the Fourier transform of the center of the cell at the EMCCD (ProEM-HS: 512BX3). This optical system maps the momentum anti-correlations of the generated photons to spatial anti-correlations in the far field. Due to the cross-Kerr effect between the pump and the probe and conjugate, the Fourier planes do not lie at the expected plane. To ensure the correct Fourier planes at the EMCCD, we place an amplitude mask (resolution test chart) in the path of the input probe and use an optical system to generate the Fourier transform of the pattern introduced by the mask at the center of the cell. This allows us to optimize the positions of the Fourier transform lenses after the cell by recovering the pattern on the EMCCD.

 Seeding the FWM to generate bright twin-beams introduces two main complications. First, since the quantum correlations in the spatial degree of freedom are contained in the relative spatial fluctuations between the probe and conjugate beams, it is necessary to extract these fluctuations from the large signal that results from the spatial profile of the input probe beam. Second, the input probe introduces classical spatial and temporal excess noise, which needs to be cancelled as much as possible to observe spatial squeezing. These issues can be resolved by acquiring two probe-conjugate images in fast succession through the kinetic mode of the EMCCD. The subtraction of these two images leads to the cancellation of the low frequency portion of the classical noise as well as the Gaussian profiles of the probe and conjugate pulses (see appendix).

To implement the acquisition of the images in rapid succession, we divide the total active sensor area, which consists of 512$\times$512 pixels (pixel size of $16\times16~\mu$m) and an additional 512$\times$512 pixel buffer region for storage, into the maximum possible number of frames,  provided that the probe and conjugate beams completely fit in each frame. The size of each frame comes out to $512\times170$ pixels, which results in a total of 6 frames. We record 100 image sequences and use the second and third frames (stored in the buffer region) for the analysis, as they have the least amount of background noise. During the image acquisition process, only the top frame is exposed. Once an image is taken in that frame, charge is serially transferred to the next frame at a speed of 300~ns/row. This limits the time separation between frames to at least $51~\mu$s. For each frame, both the pump and probe laser beams are pulsed with AOMs, resulting in pulses with $10~\mu$s and $1~\mu$s FWHM, respectively. The probe pulses are turned on $6~\mu$s after the pump pulse to avoid transient effects in the FWM. This leads to a time scale between two consecutively acquired probe pulses of $\sim60~\mu$s. The camera exposure time per frame is $12~\mu$s and the time sequence of the pump and probe pulses is synchronized with each frame (see details in appendix). While our camera has electron multiplier capabilities, we have instead used the low noise mode due to the large number of photons present in the twin-beams.

To obtain the degree of spatial quantum correlations, we look at the noise in the difference between the probe and conjugate spatial fluctuations. The spatially correlated regions between the conjugate and probe are located diametrically opposite to each other due to the phase matching condition~\cite{Lett2}.  Therefore, we first rotate the conjugate image in each frame by 180 degrees. Then, we align the corresponding spatial regions with an image registration algorithm. Once the probe and conjugate images are aligned, we crop an 80$\times$80 pixel region around the maximum intensity regions in each frame for the probe and conjugate images, and subtract these regions in two consecutive frames to obtain the spatial intensity fluctuations of the probe and conjugate. Finally, to analyze the quantum noise reduction, the spatial intensity fluctuations of the probe and conjugate are subtracted from each other. To quantify the quantum noise correlation, we introduce the noise ratio
\begin{equation}\label{NRF}
    \sigma \equiv \frac{\langle \delta^2((N_{p1}-N_{p2})-(N_{c1}-N_{c2})) \rangle}{\langle N_{p1}+N_{c1}+N_{p2}+N_{c2} \rangle},
\end{equation}
where the numerator is the spatial variance of the difference between the spatial intensity fluctuations of probe and conjugate pulses, and the denominator corresponds to the SNL. Thus, a value of $\sigma=1$ corresponds to the SNL or the noise ratio for coherent states. In Eq. (1), ($N_{p1}$, $N_{c1}$) and ($N_{p2}$, $N_{c2}$) are the matrices representing the photo-counts per pixel for the cropped regions in the probe and conjugate images for the two successive frames used for the analysis. The statistics are calculated over the pixels of the EMCCD.

Due to the finite angular spread of the Gaussian pump beam used for the FWM, there is a finite angular uncertainty in the wave vectors of the generated twin-beams.  This leads to a minimum spatial scale for the spatial quantum correlations called the coherence area~\cite{MarinoCoh}. The presence of many coherence areas in the twin-beams indicates their multi-spatial-mode nature.  The coherence area also places a constraint on the minimum detection area, as detection below this spatial scale effectively introduces losses that destroy the quantum correlations. Thus, to observe a significant level of spatial squeezing, the detection area needs to be larger than the coherence area~\cite{Marcelo,Lugiato}.

To get an estimate of the size of the coherence area, we calculate the cross-correlation function between the probe and conjugate spatial intensity fluctuations, as shown in Fig.~2(a). A section of 80$\times$80 pixels in the probe spatial intensity fluctuations is cropped and scanned over the conjugate spatial intensity fluctuations. The peak in the cross-correlation plot in Fig.~2(a) shows the presence of a correlated region between the probe and conjugate. Given a width of the cross-correlation peak of $\sim$~10$\times$10 pixels (FWHM), the size of the coherence area is of $\sim$~7$\times$7 pixels (FWHM) after deconvolving. The low value of the cross-correlation peak is due to our detection area (single pixel) being smaller than the coherence area. We have also performed the same analysis for two coherent state pulses and, as expected, no correlation is observed (Fig.~2(b)).

\begin{figure}[hbt]
\centering
\includegraphics{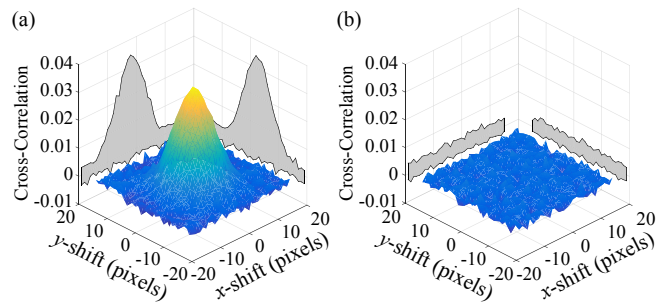}
\caption{Cross-correlation plots between (a) probe and conjugate pulses and (b) two coherent state pulses.}
\label{fig:Fig2}
\end{figure}

\begin{figure}[hbt]
\centering
\includegraphics{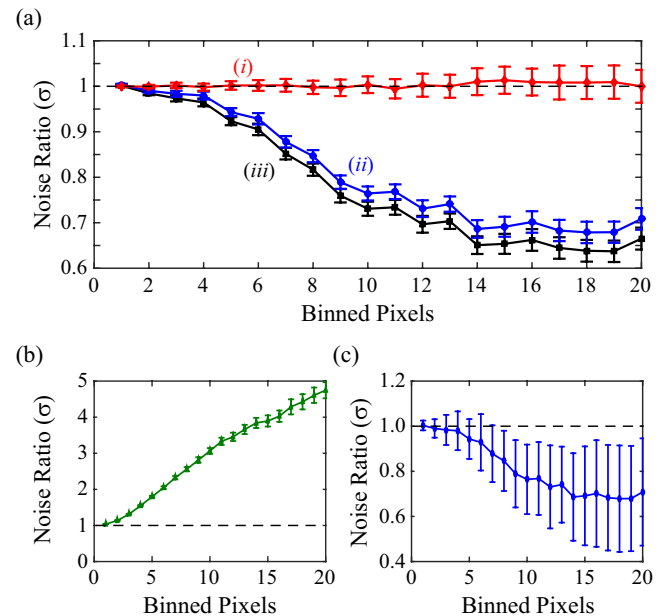}
\caption{Experimentally observed noise ratio for twin-beams and coherent state pulses at different binnings. Number of binned pixels given along x-axes represent the number of pixels used along each side of the square binning region. (a) Noise ratios for: ($i$) coherent state pulses and ($ii$, $iii$) pulsed twin-beams without and with background correction, respectively. For the twin-beam traces conjugate images are rotated 180 degrees with respect to the probe images. (b) Noise ratio without rotation of the conjugate images. Error bars in (a, b) represent the standard deviation of the mean noise ratio over 100 shots. (c) Noise ratio of trace ($ii$) in (a) with error bars representing the standard deviation of noise ratio over the 100 shots.}
\label{fig:Fig3}
\end{figure}

To see the dependence of the size of the detection area with respect to the coherence area on the spatial squeezing, we perform the noise analysis while grouping different numbers of pixels in the images by binning them into square regions of $n\times n$ pixels in the computer and defining super-pixels. To keep the total number of super-pixels as an integer, we have used slightly different analysis regions for different binnings. We calculate the noise ratio given in Eq.~(1) for different pixel binnings and plot it in Fig.~3 as a function of the number of pixels grouped ($n$) along each side of the square binning region. As can be seen in Fig.~3(a), with 180 degrees rotation of the conjugate images before aligning correlated regions of probe and conjugate, the noise ratio for the twin-beams (trace $ii$) falls below one, i.e. below the SNL. With increasing binning, the noise ratio is further and further reduced until it saturates at higher binning at a level significantly below the SNL. Note that $\sigma$ starts to saturate at a binning slightly higher than the size of coherence area of $\sim$7$\times$7 pixels. We also calculate the noise ratio with background noise correction as done in  Ref.~\cite{shot3} (see appendix).  As can be seen in Fig~3(a) (trace $iii$), a slightly larger noise reduction is obtained with background noise correction. The minimum noise ratios obtained without and with background correction are 0.67$\pm$0.02 and 0.63$\pm$0.02 respectively; which correspond to~-1.74$\pm$0.13~dB and -2.00$\pm$0.14~dB of spatial squeezing respectively. The error bars represent the standard deviation of the mean noise ratio. The total photo-counts in the analysis region of the probe pulse are $\sim$ 1.8 $\times$ 10$^8$. As a test of the data acquisition and data analysis, we used a similar setup to verify the SNL for coherent state pulses with the same number of photons as the pulsed twin-beams. The experimental result of the noise ratio for coherent state pulses is shown in trace~$i$ of Fig.~3(a), and, as expected $\sigma\sim 1$, i.e. the SNL.

In order to study the effect of the Gaussian profile of the twin-beams on the calculated noise ratios, we also performed the data analysis by normalizing the intensities of the probe and conjugate pulses in the analysis region such that the beam profiles have a uniform intensity before performing the noise analysis. This results in the same weighting for every pixel in the analysis region. We find that there is no significant change from the results shown in Fig.~3(a). This gives an indication that the degree of spatial squeezing is uniform across the transverse profile of the bright pulsed twin-beams in the analysis region.

To illustrate the anti-correlations in momentum, we calculate the noise ratio with non-rotated conjugate images, as shown in Fig.~3(b). In this case the difference noise is calculated between uncorrelated spatial regions of the probe and conjugate pulses; thus the noise adds in quadrature and leads to excess noise. While the error bars in Figs.~3(a) and (b) present the standard deviation of the mean, we also show the noise ratio with error bars representing the standard deviation over the 100 shots in Fig.~3(c). One can notice that the size of the error bars increases with increasing binning. As the number of binned pixels increases, the number of super-pixels over which $\sigma$ is calculated, is reduced. Thus the error bars are limited by the statistics. However, it is worth noting that even with these limited statistics, we can obtain spatial squeezing in most of the single shot measurements.

\begin{figure}[hbt]
\centering
\includegraphics{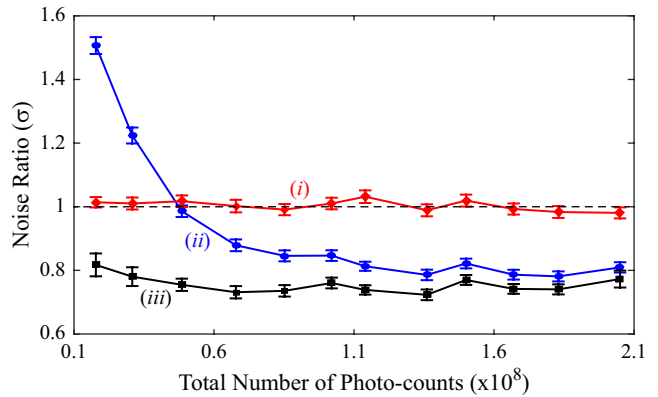}
\caption{Experimentally observed noise ratio for twin-beams and coherent state pulses as a function of total photo-counts in the 1~$\mu$s probe beam pulse. The different traces correspond to: ($i$) coherent state pulses, ($ii$) pulsed twin-beams without background subtraction and ($iii$) pulsed twin-beams with background subtraction. Error bars represent standard deviation of the mean noise ratio over the 100 acquired shots.}
\label{fig:Fig4}
\end{figure}

We also study the dependence of the noise ratio on the number of photons in the twin-beams with and without background noise correction, as shown in Fig.~4. We consider a maximum of $\sim 2\times10^8$ total number of photons in the probe pulse, limited by the saturation of the EMCCD and not by the source. For the purpose of statistics, we have performed the analysis for a binning of 10$\times$10 pixels. As can be seen in Fig.~4~(trace $ii$), the noise ratio is below the SNL for a wide range of intensities of the pulsed twin-beams. However, for a total number of photo-counts lower than $\sim5\times10^7$, background noise dominates and results in excess noise. For these cases, background noise correction becomes relevant and spatial squeezing can be recovered after taking the background noise into account, as shown in Fig.~4~(trace $iii$). Furthermore, for coherent state pulses, we obtain the SNL independent of the number of photo-counts, as shown in Fig.~4~(trace $i$).

There are two main factors limiting the minimum noise reduction or spatial squeezing that can be measured. First, the losses through the system degrade the squeezing. If $\eta$ is the total optical transmission, including detection efficiency, then the minimum noise ratio is limited to $1-\eta$. We have 94.5$\%$ transmission through the optical system after the cell and 70$\%$ quantum efficiency of the EMCCD at -50$^{\circ}$C. With these losses, the minimum noise ratio that can be measured is limited to 0.34, or -4.68~dB. Second, unlike the noise reduction in SPDC where $\sigma$ can, in principle, go to zero in the absence of losses, with FWM we are limited by the shot noise introduced by the input probe beam. In the ideal case of no losses, this limits the noise ratio to $1/(2G-1)$, where $G$ is the FWM gain (4.5 in the current experiment). When working with higher gains, the contribution of this noise is minimized. This has made it possible to obtain $\sim$9~dB of squeezing in the time domain with this source ~\cite{Lett1,Glorieux}. Besides these factors, the main source of excess noise is due to scattered pump photons, which can be minimized by using an atomic line filter~\cite{Mitchell}. This would make it possible to operate at lower number of photons.

The use of bright twin-beams does offer unique advantages. First, it is possible to control the number of photons independent of the level of the squeezing. Second, seeding the FWM process leads to spatially localized twin-beams, which makes it possible to work in the high gain regime and avoid some of the problems that have limited the level of the squeezing and the number of photons in bright squeezed vacuum states~\cite{Leuchs1,Leuchs2}.

In conclusion, we have experimentally shown spatial squeezing with bright twin-beams of light. The use of an EMCCD allows us to obtain a measure of the size of the coherence area through cross-correlation measurements of the spatial fluctuations. Furthermore, with detection area greater than the coherence area, we observed a quantum noise reduction of around 2~dB with respect to the SNL. A consistent level of spatial squeezing is observed for a range of total photo-counts in the pulsed twin-beams.

Spatial squeezing in the macroscopic domain opens a new avenue for quantum imaging and quantum metrology, as it allows for a significant enhancement in the signal-to-noise ratio in a single shot. This makes bright twin-beams ideal for real-time imaging applications. Furthermore, the narrowband photons generated with the FWM process are ideal for interacting with atomic ensembles in quantum information protocols.

This work was supported by the W. M. Keck Foundation.

\section{Appendix}
Here, we give the details of the data acquisition and the data analysis for the results presented in the main text.

\subsection{Data Acquisition}
\indent Figure~5(a) shows the time sequence of the pump and the input probe pulses used to capture twin-beam images in two consecutive EMCCD frames. Each frame contains an image of the probe and an image of the conjugate, as shown in Fig.~5(b). The time sequence of the pump and input probe pulses is synchronized with the acquisition time of each frame. This is done by sending an external trigger pulse from the camera to two arbitrary function generators that drive the AOMs used to generate the pulses. This leads to a time scales between two consecutive input probe pulses of $\sim60~\mu$s.\\

\begin{figure}[hbt]
\centering
\includegraphics[width=\columnwidth]{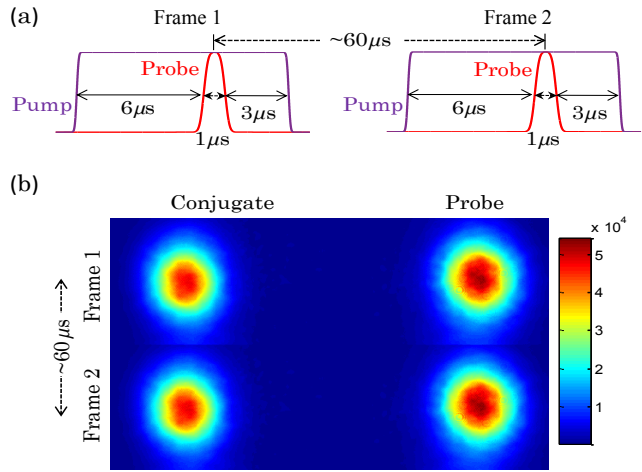}
\caption{(a) Timing sequence of the pump and input probe pulses used to acquire probe and conjugate images in two consecutive frames of the EMCCD. (b) Two consecutive frames (frame size $170\times512$ pixels) acquired by the EMCCD camera. Each frame contains an image of the probe (right) and an image of the conjugate (left). The time interval between these two frames is $\sim60~\mu$s.}
\label{fig:Fig5}
\end{figure}

\subsection{Data Analysis}
\indent The technique used for the analysis of the spatial quantum correlations is analogous to the technique that is routinely implemented in the time domain with a spectrum analyzer (SA) when looking for squeezing with bright twin-beams. In particular, when performing measurements in the time domain, the SA allows us to filter out the ``DC'' or bright portion of the beams and the squeezing is characterized at an analysis frequency different from ``DC''. That is, the quantum correlations are characterized through a study of the correlations in the temporal fluctuations of the field \cite{BachorBook}.

We perform the corresponding analysis in the spatial domain by subtracting the images in two consecutive frames (shown in Fig.~5(b)). This allows us to ``filter'' out the spatial ``DC'' part of the images (or spatial profile) to extract the  spatial fluctuations and measure the spatial quantum correlations. It is worth noting that image processing would also allow us to filter out the spatial ``DC'' portion. However, the exact technique (cutoff frequency, filter type, etc.) that is used in the image processing can lead to unwanted artifacts, such as filtering out the actual spatial fluctuations that give rise to the quantum correlations. We thus opted to implement the two frames subtraction to avoid these artifacts.

After the two frames subtraction, only the spatial intensity fluctuations in each beam remain. We can then use the noise ratio ($\sigma$), defined in Eq.~(1) in the main text, to characterize the degree of relative spatial correlations between the probe and conjugate. This leads to a noise ratio smaller than one i.e. below the SNL, as shown in Figs.~3 and 4 in the main text.

It is important to note that the time between the two frames used for the analysis is significantly larger than the inverse bandwidth of the process i.e. $\sim 1/$(15~MHz) or $\sim 66$~ns \cite{Lett6,Howell}, which means that the twin-beam pulses in two consecutive frames are uncorrelated from each other. As a result, their noise adds in quadrature and the noise ratio obtained through Eq.~(1) in the main text characterizes the spatial correlations.

\subsection{Background Noise Correction}
As shown in Fig.~4 in the main text, the main source of excess noise is due to the scattered pump photons. To take this noise into account, we capture the background scattered pump noise in each EMCCD frame by turning off the input probe beam after each probe-conjugate image acquisition. For the background noise analysis, we first locate the corresponding probe and conjugate image positions in the background noise images and crop the regions equal to the probe and conjugate analysis regions in the background images. We perform the same binning for the cropped background images as we do for the probe and conjugate analysis regions. Finally the background noise subtraction from the noise ratio is performed as follows

 \begin{widetext}
 \begin{equation*}
   \sigma_{B} \equiv \frac{\langle \delta^2((N_{p1}-N_{p2})-(N_{c1}-N_{c2}))\rangle-\langle \delta^2((N_{pb1}-N_{pb2})-(N_{cb1}-N_{cb2}))\rangle}{\langle N_{p1}+N_{c1}+N_{p2}+N_{c2} \rangle-\langle N_{pb1}+N_{bc1}+N_{bp2}+N_{bc2} \rangle},
    \end{equation*}
 \end{widetext}
where ($N_{pb1}$, $N_{cb1}$) and ($N_{pb2}$, $N_{cb2}$) are the matrices representing the photo-counts per super-pixel for the background noise photo-counts in two consecutive frames.
The background images contain $\sim 10^6$ total number of scattered pump photons in the analysis region.

\end{document}